\documentclass{jcmlatex}
\usepackage{amssymb}
\usepackage{graphicx}

\def\JCMvol{xx}
\def\JCMno{x}
\def\JCMyear{200x}
\def\JCMreceived{xxx}
\def\JCMrevised{xxx}
\def\JCMaccepted{xxx}

\begin{document}

\markboth{L. FERN\'ANDEZ-JAMBRINA}
         {Rational developable surfaces}

\title{CHARACTERISATION OF RATIONAL AND NURBS DEVELOPABLE SURFACES IN COMPUTER 
AIDED DESIGN}%

\author{Leonardo Fern\'andez-Jambrina
\thanks{ETSI Navales, Universidad Polit\'ecnica de Madrid, 
28040-Madrid, Spain\\
  Email: leonardo.fernandez@upm.es
          } 
    }

\maketitle

\begin{abstract}
In this paper we provide a characterisation of rational developable
surfaces in terms of the blossoms of the bounding curves and three
rational functions $\Lambda$, $M$, $\nu$.  Properties of developable
surfaces are revised in this framework.  In particular, a closed
algebraic formula for the edge of regression of the surface is
obtained in terms of the functions $\Lambda$, $M$, $\nu$, which are
closely related to the ones that appear in the standard decomposition
of the derivative of the parametrisation of one of the bounding curves
in terms of the director vector of the rulings and its derivative.  It
is also shown that all rational developable surfaces can be described
as the set of developable surfaces which can be constructed with a
constant $\Lambda$, $M$, $\nu$ .  The results are readily extended to
rational spline developable surfaces.
\end{abstract}



\section{Introduction}

Ruled surfaces are useful since the simplest way to interpolate a 
surface patch between two given curves is to link them with straight 
segments. Ruled surfaces have non-positive Gaussian curvature, since 
in general the straight lines that they contain are not lines of 
curvature of the surface. In developable surfaces the 
straight lines are one of the families of lines of curvature and 
hence these surfaces have null Gaussian curvature. 

Mathematically, this means that developable surfaces are isometric to 
the plane. Extrinsic, but not intrinsic, curvature arises from the way these surfaces are embedded 
in space. Since distances, areas and angles are conserved on 
embedding the surfaces in space, this means that developable surfaces 
are plane patches which have been folded or cut, but not deformed in 
any other fashion. Rolling pieces of planes in 
cones and cylinders are the most obvious ways of achieving this, but 
there are more general and less intuitive ways.

For such reason developable surfaces are valuable for applications in
industry.  Developable surfaces model the way the pages of a book are
folded \cite{bartoli}, the forms of facades in architecture
\cite{architecture} or the shapes adopted by garments \cite{rose} with
plane patterns.  They are also useful in industries related to
building with sheets of steel or wood, such as naval industry
\cite{kilgore, chalfant, arribas}, or even automobile industry
\cite{frey}.  In the case of steel this means that parts of the hull
of a ship can be modeled with developable surfaces and can be produced
by folding machines without application of heat, reducing costs and
modifications of the metallic structure.

Since Gaussian curvature is the quotient of the determinants of the 
fundamental forms of the surface, its calculation involves non-linear 
combinations of the derivatives of the parametrisation of the 
surface. If we think of applications to Computer Aided Geometric 
Design, this translates into non-linear expressions in terms of the 
control points and weights of the surface. An extensive review on 
this issue appears in \cite{computational-line}.

In the case of rational surfaces, conditions for null Gaussian 
curvature can be solved for low degrees \cite{lang}, but there are 
other approaches to this issue. For instance, restriction to boundary 
curves on parallel planes simplifies the problem  
\cite{aumann0,maekawa}.

A geometrically appealing approach relies on projective geometry.  In
dual space points are planes in space.  Since developable surfaces can
be viewed as envelopes of one-parametric families of planes, dual
space appears as a natural framework \cite{ravani, pottmann-farin,
wallner}, though the actual control points lie on ordinary space. 
In \cite{origami} the null Gaussian curvature condition is
written in terms of quadratic equations in order to devise a
constraint useful for interactive modeling.

Also within the NURBS framework, the properties of the de
Casteljau algorithm have been explored for constructing developable
surfaces \cite{sequin}.  In \cite{aumann, aumann1} B\'ezier
developable patches are constructed by applying affine transformations
to the first cell of the control net of the patch.  It is shown in
\cite{leonardo-bezier} that this construction produces all B\'ezier
developable surfaces with a polynomial edge of regression.  This
construction has been extended to spline developable surfaces
\cite{leonardo-developable, leonardo-elevation} and to B\'ezier
triangular surfaces \cite{leonardo-triangle}.

Another interesting approach for designing approximately developable
surfaces ia based on the use of the convex hulls of the boundary
\cite{rose}. Other approximations may be found in \cite{leopoldseder, 
arribas}.

Most recently \cite{rabinovich} presents a new approach grounded on
the characterisation of developable surfaces as surfaces parametrised
by orthogonal sets of geodesics. \cite{stein} suggests producing 
developable triangular meshes in order to design developable 
surfaces. \cite{leonardo-poly} contructs developable patches bounded 
by two curves, reparametrising one of the curves.

Since the standard of Computer Aided Design (CAD) is based on the use
of rational B-spline curves and surfaces, one would require a
description of rational developable surfaces within this framework.
That is, involving the elements that are used in design for defining
curves and surfaces. such as control points, knots and weights.  It would be
interesting hence to extend Aumann's approach \cite{aumann,leonardo-bezier}
from polynomial to rational developable surfaces in order to comply
with the whole NURBS framework. The main advantage of this approach 
is the use of the elements which are used in CAD applications.

This paper is organised as follows.  Section 2 is devoted to an
introduction to developable surfaces as envelopes of families of
planes and their classification in terms of their edge of regression.
Section 3 provides a characterization of rational developable surfaces
based on the de Casteljau algorithm, in terms of three rational
functions, $\Lambda$, $M$, $\sigma$.  Section~4 discusses a useful way
of parametrising rational ruled surfaces, which allows an
interpretation for $\Lambda$, $M$, $\sigma$. Function $\sigma$ can become
trivial by a suitable choice of a global factor for the parametrisation.  The
main properties of rational developable surfaces in our framework are
described in Section~5.  The edge of regression of rational
developable surfaces is calculated in closed form in Section~6.  It is
shown that it is a rational curve of degree $n+1$ for rational
developable patches with bounding curves of degree $n$ in the case of
constant $\Lambda$, $M$, $\sigma$.  The converse is also true, that is, every
rational developable surface admits surface patches with constant
$\Lambda$, $M$, $\sigma$.  This suggests that one may start with constant
$\Lambda$, $M$, $\sigma$ patches and modify the length of the rulings
afterwards to adapt them to one's purposes.  The construction of
constant $\Lambda$, $M$, $\sigma=1$ patches is derived in Section~7. 
Examples are shown at the end of the paper.

\section{Developable surfaces}
\begin{figure}
\begin{center}
    \includegraphics[height=0.2\textheight]{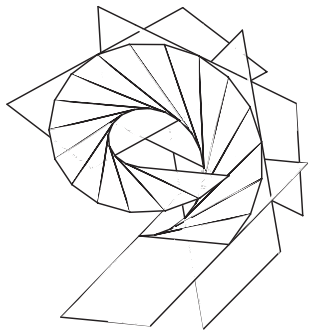}
\end{center}
\caption{Developable surface as envelope of a family of planes}
\label{heli}\end{figure}

Developable surfaces may be viewed as envelopes of uniparametric 
families of planes \cite{struik},\[\mathbf{a}(\lambda)\cdot \mathbf{x}+ 
b(\lambda)=0,\]where $\mathbf{a}(\lambda)$ is a normal vector to the 
plane assigned to the parameter $\lambda$. The envelope of this 
family (see Fig. \ref{heli}), if it exists, is a surface fulfilling the equations
\[\mathbf{a}(\lambda)\cdot \mathbf{x}+ 
b(\lambda)=0,\qquad \mathbf{a}'(\lambda)\cdot \mathbf{x}+ 
b'(\lambda)=0.\]

For each value of $\lambda$ the former equations provide the
intersection of the enveloping surface with the corresponding plane.
Since these intersections are straight lines (called \emph{rulings}), the
enveloping surface is ruled.  Furthermore, since the enveloping
surface is tangent to each member of the family along their
intersection, the tangent plane to the developable surface along each 
ruling is the same for all points on the ruling. 

Hence, if we parametrise a developable surface as a ruled surface
patch between two smooth parametrised curves $c(t)$, $d(t)$,
\[b(t,v)=(1-v)c(t)+vd(t),\]
the constant tangent plane requirement can be expressed as a 
coplanarity condition between the velocities of the parametrised 
curves, $c'(t)$, $d'(t)$ and the vector connecting points with the 
same parameter $t$ (cfr. for instance \cite{leonardo-developable}),
\begin{equation}\label{copla}
\left(d(t)-c(t)\right)\cdot \left(c'(t)\times d'(t)\right)=0.
\end{equation}

This condition fully characterises developable surfaces since every 
surface satisfying it is necessarily the envelope of the family of 
its tangent planes.

\begin{figure}
\begin{center}
    \includegraphics[height=0.2\textheight]{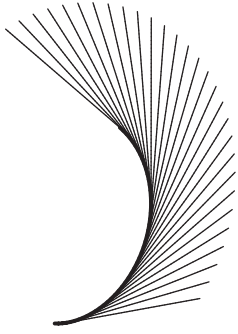}
\end{center}
\caption{Edge of regression as envelope of the family of rulings}
\label{edge}\end{figure}

We may also consider the rulings of the developable surface as a 
uniparametric family of curves. The envelope of this family of 
straight lines, if it exists, 
\[\mathbf{a}(\lambda)\cdot \mathbf{x}+ 
b(\lambda)=0,\qquad \mathbf{a}'(\lambda)\cdot \mathbf{x}+ 
b'(\lambda)=0,\qquad \mathbf{a}''(\lambda)\cdot \mathbf{x}+ 
b''(\lambda)=0,\]
is a curve called \emph{edge of regression} of the developable surface,
which is tangent to every ruling at a point $\gamma(\lambda)$ (see 
Fig. \ref{edge}). This 
assignment to each value of the parameter $\lambda$ to a point on the 
developable surface serves as a parametrization of the edge of 
regression. 

Since we may parametrise the developable surface as
$g(t,\lambda)=\gamma(\lambda)+ t\,\gamma'(\lambda)$, the singular
points of the surface satisfy \[0=\frac{\partial
g(t,\lambda)}{\partial t}\times \frac{\partial
g(t,\lambda)}{\partial\lambda} =
\gamma'(\lambda)\times\left(\gamma'(\lambda)+
t\,\gamma''(\lambda)\right) = t\gamma'(\lambda)\times
\gamma''(\lambda),\]
and we notice that the edge of regression, if it exists, is part of the 
set of singular points of the developable surface.

Therefore, it is important to keep track of the edge of regression 
when modelling in order to avoid the appearance of undesired 
singularities.

\begin{figure}
\begin{center}
    \includegraphics[height=0.15\textheight]{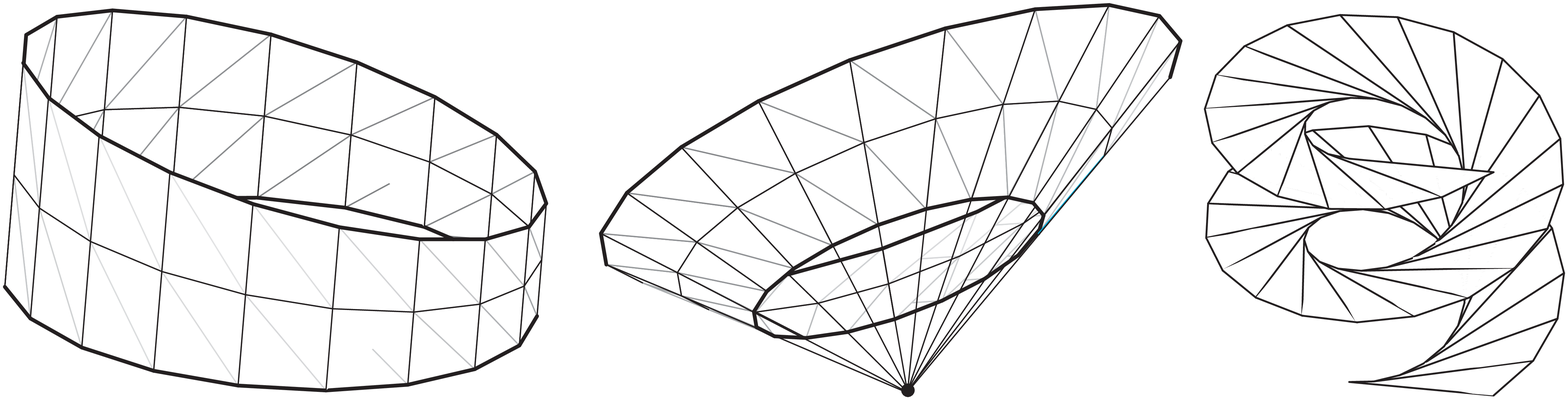}
\end{center}
\caption{Classes of developable surfaces}
\label{clase}\end{figure}

Leaving out plane surfaces, which are trivially developable, we may
classify developable surfaces according to their edge of regression 
into three families (see Fig. \ref{clase}):

\begin{itemize}
\item  Cylindrical surfaces: developable surfaces with no edge of 
regression at all. All rulings are parallel.

\item Conical surfaces: developable surfaces for which the edge 
of regression is degenerate and reduces to a point, the vertex of 
the cone, where all rulings meet.

\item Tangent surfaces: the generic case of a developable surface 
with an edge of regression which is an actual curve.
\end{itemize}

The first two families are well understood within the NURBS formalism,
are easy to construct and have therefore been used extensively in 
geometric modelling on resorting to developable surfaces. What it is 
aimed here is to produce a description of developable surfaces which 
may include the generic case of tangent surfaces.

\section{Rational developable surfaces}

In order to describe rational developable surfaces we start by
considering a ruled surface interpolated between two rational curves
of degree $n$, $c(t)$, $d(t)$, defined by their respective control
polygons, $\{c_{0},\ldots,c_{n}\}$, $\{d_{0},\ldots,d_{n}\}$, and sets
of weights $\{w_{0},\ldots,w_{n}\}$,
$\{\omega_{0},\ldots,\omega_{n}\}$,
\[c(t)=\frac{\displaystyle\sum_{i=0}^n w_{i}c_{i}B^n_{i}(t)}
{\displaystyle\sum_{i=0}^n w_{i}B^n_{i}(t)},\qquad
d(t)=\frac{\displaystyle\sum_{i=0}^n \omega_{i}d_{i}B^n_{i}(t)}
{\displaystyle\sum_{i=0}^n \omega_{i}B^n_{i}(t)},\]
in terms of the Bernstein polynomials of degree $n$, or the de 
Casteljau algorithm \cite{farin},
\begin{eqnarray}\label{castel}
   w^{r}_{i}(t)&=&(1-t) w^{r-1}_{i}(t)+t w^{r-1}_{i+1}(t),\quad
   i=0,\ldots, n-r,\quad r=1,\ldots, n, \nonumber\\ 
   c^{r}_{i}(t)&=&(1-t) \frac{w^{r-1}_{i}(t)}{w^{r}_{i}(t)}c^{r-1}_{i}(t)+
   t \frac{w^{r-1}_{i+1}}{w^{r}_{i}(t)}c^{r-1}_{i+1}(t), 
   \nonumber\\
   c(t)&:=&c^{n}_{0}(t)=(1-t)\frac{w^{n-1}_{0}(t)}{w^{n}_{0}(t)}c^{n-1}_{0}(t)+
   t \frac{w^{n-1}_{1}}{w^{n}_{0}(t)}c^{n-1}_{1}(t)
\end{eqnarray} where $w^{0}_{i}=w_{i}$ and $c^{0}_{i}=c_{i}$.

From the derivatives of both numerator, $p(t)=w(t)c(t)$, and 
denominator of a rational curve $c(t)$,
\begin{eqnarray}p'(t)&=&n\left(w^{n-1}_{1}(t)c^{n-1}_{1}(t)-w^{n-1}_{0}(t)c^{n-1}_{0}(t)\right),
\nonumber\\
w'(t)&=&n\left(w^{n-1}_{1}(t)-w^{n-1}_{0}(t)\right),\end{eqnarray}
we get the derivative of a rational curve $c(t)$ \cite{floater},
\[c'(t)=\frac{p'(t)-w'(t)c(t)}{w(t)}=\frac{nw_{0}^{n-1}(t)w_{1}^{n-1}(t)}
{w_{0}^{n}(t)^2}\left(c_1^{n-1}(t)-c_0^{n-1}(t)\right),\]
as a difference between the two last-but-one points in the de Casteljau 
algorithm.

Hence we have seen that the vectors $c'(t)$, $d'(t)$, $d(t)-c(t)$ are
barycentric combinations of the points $c^{n-1}_{0}(t)$,
$c^{n-1}_{1}(t)$, $d^{n-1}_{0}(t)$, $d^{n-1}_{1}(t)$.  Therefore
$c'(t)$, $d'(t)$, $d(t)-c(t)$ are coplanary if and only if
$c^{n-1}_{0}(t)$, $c^{n-1}_{1}(t)$, $d^{n-1}_{0}(t)$, $d^{n-1}_{1}(t)$
lie on a plane and the developability condition (\ref{copla}) for a
rational ruled surface may be restated in terms of these:

\begin{proposition}The ruled surface interpolating between two rational curves of 
degree $n$, defined by their respective control
polygons, $\{c_{0},\ldots,c_{n}\}$, $\{d_{0},\ldots,d_{n}\}$, and sets
of weights $\{w_{0},\ldots,w_{n}\}$, 
$\{\omega_{0},\ldots,\omega_{n}\}$, is developable if and only if the points $c^{n-1}_{0}(t)$, 
$c^{n-1}_{1}(t)$, $d^{n-1}_{0}(t)$, $d^{n-1}_{1}(t)$ are 
coplanary.\end{proposition}

In order to avoid denominators, we may rewrite this result in terms 
of vectors in $\mathbb{R}^4$,
$\mathbf{p}_0^{n-1}(t)$, $\mathbf{p}_1^{n-1}(t)$, 
$\mathbf{q}_0^{n-1}(t)$, $\mathbf{q}_1^{n-1}(t)$,
\begin{eqnarray*}\mathbf{p}^{n-1}_{i}(t)&=&\left(w_{i}^{n-1}(t),w_{i}^{n-1}(t)c_{i}^{n-1}(t)\right),\\
\mathbf{q}^{n-1}_{i}(t)&=&\left(\omega_{i}^{n-1}(t),\omega_{i}^{n-1}(t)d_{i}^{n-1}(t)
\right).\end{eqnarray*}

\begin{figure}[h]\begin{center}
\includegraphics[height=0.25\textheight]{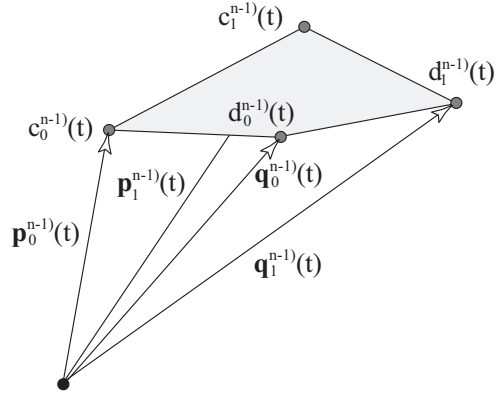}
\caption{Points $c^{n-1}_{0}(t)$, 
$c^{n-1}_{1}(t)$, $c^{n-1}_{0}(t)$, $c^{n-1}_{1}(t)$ are coplanary iff vectors $\mathbf{p}^{n-1}_{0}(t)$, 
$\mathbf{p}^{n-1}_{1}(t)$, $\mathbf{q}^{n-1}_{0}(t)$, 
$\mathbf{q}^{n-1}_{1}(t)$ lie on a 3-space}
\label{coplane}\end{center}
\end{figure}

The condition of coplanarity for the points in affine space
(Fig.~\ref{coplane}) becomes a condition of linear dependence for the
corresponding vectors in $\mathbb{R}^{4}$:

\begin{corollary}The ruled surface interpolating between two rational curves of 
degree $n$, defined by their respective control
polygons, $\{c_{0},\ldots,c_{n}\}$, $\{d_{0},\ldots,d_{n}\}$, and sets
of weights $\{w_{0},\ldots,w_{n}\}$, 
$\{\omega_{0},\ldots,\omega_{n}\}$, is developable if and only if the 
vectors $\mathbf{p}^{n-1}_{0}(t)$, 
$\mathbf{p}^{n-1}_{1}(t)$, $\mathbf{q}^{n-1}_{0}(t)$, $\mathbf{q}^{n-1}_{1}(t)$ are 
linearly dependent.\end{corollary}

That is, there exist rational coefficients $\lambda_{0}(t)$, $\lambda_{1}(t)$, 
$\mu_{0}(t)$, $\mu_{1}(t)$, such that
\[\lambda_{0}(t)\mathbf{p}^{n-1}_{0}(t)+\lambda_{1}(t)\mathbf{p}^{n-1}_{1}(t)=
\mu_{0}(t)\mathbf{q}^{n-1}_{0}(t)+\mu_{1}(t)\mathbf{q}^{n-1}_{1}(t).\]

The coefficients in the equation are defined up to a multiplicative 
factor and we may divide it by $\lambda_{0}(t)+\lambda_{1}(t)$ and 
define 
$\Lambda(t)=\lambda_{1}(t)/\left(\lambda_{0}(t)+\lambda_{1}(t)\right)$, 
$M(t)=\mu_{1}(t)/\left(\mu_{0}(t)+\mu_{1}(t)\right)$, 
$\sigma(t)=\left(\mu_{0}(t)+\mu_{1}(t)\right)/\left(\lambda_{0}(t)+\lambda_{1}(t)\right)$,
\begin{equation}\left(1-\Lambda(t)\right)\mathbf{p}^{n-1}_{0}(t)+\Lambda(t)\mathbf{p}^{n-1}_{1}(t)=
\sigma(t)\left(\left(1-M(t)\right)\mathbf{q}_0^{n-1}(t)+M(t)
\mathbf{q}_1^{n-1}(t)\right).\end{equation}

This way of writing the linear combination excludes the case of 
$\lambda_{1}=-\lambda_{0}$. 
However, it does not hinder our goal of coping with the generic case.

We may gain insight into this result by rewriting it in terms of 
blossoms,
\begin{eqnarray}
\mathbf{p}^{1}_{i}[t_{1}]&:=&\mathbf{p}^{1}_{i}(t_{1})=(1-t_{1}) \mathbf{p}_{i}+t_{1} \mathbf{p}_{i+1},\quad i=0,\ldots, 
n-1,\nonumber\\
\mathbf{p}^{r}_{i}[t_{1},\ldots,t_{r}]&:=&(1-t_{r}) 
\mathbf{p}_{i}^{r-1}[t_{1},\ldots,t_{r-1}]+t_{r} \mathbf{p}_{i+1}^{r-1}[t_{1},\ldots,t_{r-1}]
,\nonumber\\
\mathbf{p}[t_{1},\ldots,t_{n}]&:=&\mathbf{p}^{n}_{0}[t_{1},\ldots,t_{n}],\quad i=0,\ldots, 
n-r,\quad r=1,\ldots,n,
\end{eqnarray}
\[\mathbf{p}_0^{n-1}(t)=\mathbf{p}[t^{<n-1>},0],\quad
\mathbf{p}_1^{n-1}(t)=\mathbf{p}[t^{<n-1>},1],\qquad  t^{<a>}:=\underbrace{t,\ldots,t}_{a}
,  \]
since linear combinations can be written in a rather compact form, taking into account that 
blossoms are multi-affine,
\begin{equation}
\mathbf{p}[t^{<n-1>},\Lambda(t)]=\sigma(t)\mathbf{q}[t^{<n-1>},M(t)].
\end{equation}


We have therefore characterised developability of a generic rational ruled 
surface in terms of the polar forms of the bounding curves of the 
patch:
\begin{theorem}\label{ratheo}
Two rational curves $c(t)$, $d(t)$ of degree $n$ with control polygons
$\{c_{0},\ldots,c_{n}\}$, $\{d_{0},\ldots,d_{n}\}$ and weights
$\{w_{0},\ldots,w_{n}\}$, $\{\omega_{0},\ldots,\omega_{n}\}$ define a generic developable surface
 if and only if there exist rational functions $\Lambda(t)$, $M(t)$, 
 $\sigma(t)$ such that the blossoms of the curves in $\mathbb{R}^4$ are related by
\[\mathbf{p}[t^{<n-1>},\Lambda(t)]=\sigma(t)\mathbf{q}[t^{<n-1>},M(t)].\] 
\end{theorem}

This expression is valid not just for rational B\'ezier curves, but
also for rational spline curves, as it is done in
\cite{leonardo-developable} from B\'ezier to splines curves. The only 
difference between the B\'ezier and the spline cases is the 
expression of the blossom, which depends on the list of knots for 
splines.
\begin{eqnarray*}\label{deboorb}
\mathbf{p}^{1)}_i[t_1]&:=&\mathbf{p}[u_{i+1},\ldots,u_{i+n-1},t_1]\;,\nonumber\\
&=&\frac{u_{i+n}-t_1}{u_{i+n}-u_{i}}\mathbf{p}_{i}+\frac{t_1-u_{i}}{u_{i+n}-u_{i}}\mathbf{p}_{i+1}
\;,\qquad i=0,\ldots,n-1\;,\nonumber\\
\mathbf{p}^{r)}_i[t_1,\ldots,t_r]&:=&\mathbf{p}[u_{i+r},\ldots,u_{i+n-1},t_1,\ldots,t_r]\nonumber
\\
&=&\frac{u_{i+n}-t_r}{u_{i+n}-u_{i+r-1}}\mathbf{p}^{r-1)}_{i}[t_1,\ldots,t_{r-1}]
\nonumber\\ &+&
\frac{t_r-u_{i+r-1}}{u_{i+n}-u_{i+r-1}}\mathbf{p}^{r-1)}_{i+1}[t_1,\ldots,t_{r-1}]\;,
\\
i&=&0,\ldots,n-r,\ r=1,\ldots,n\;,\nonumber\\
\mathbf{p}[t_1,\ldots,t_n]&:=&\mathbf{p}^{n)}_0[t_1,\ldots,t_{n}]
=\frac{u_{n}-t_n}{u_{n}-u_{n-1}}\mathbf{p}^{n-1)}_{0}
[t_1,\ldots,t_{n-1}]\nonumber\\&+&\frac{t_n-u_{n-1}}{u_{n}-u_{n-1}}\mathbf{p}^{n-1)}_{1}
[t_1,\ldots,t_{n-1}]\;,\\\mathbf{p}(u)&=&\mathbf{p}[u,\ldots,u]:=\mathbf{p}[u^{<n>}]\;,
\end{eqnarray*}
\begin{theorem}\label{sptheo}
Two rational spline curves $c(t)$, $d(t)$ of degree $n$ and $N$
pieces, with respective control polygons $\{c_{0},\ldots,c_{n+N-1}\}$,
$\{d_{0},\ldots,d_{n+N-1}\}$, weights $\{w_{0},\ldots,w_{n+N-1}\}$,
$\{\omega_{0},\ldots,\omega_{n+N-1}\}$ and common list of knots 
$\{t_{0},\ldots,t_{2n+N-2}\}$ define a generic developable
surface if and only if there exist rational functions
$\Lambda(t)$, $M(t)$, $\sigma(t)$ such that the blossoms of the curves
in $\mathbb{R}^4$ are related by
\[\mathbf{p}[t^{<n-1>},\Lambda(t)]=\sigma(t)\mathbf{q}[t^{<n-1>},M(t)].\] 
\end{theorem}

We focus on rational developable surfaces from now on, since the 
extension to rational splines is seen to be straightforward.

\section{Reparametrisation of rational ruled surfaces}

There are two convenient alternative ways of parametrising rational ruled 
surfaces. If we have two rational curves of degree $n$ parametrised 
as $c(t)=p(t)/w(t)$, $d(t)=q(t)/\omega(t)$, the standard parametrisation would be
\[b(t,v)=(1-v)c(t)+vd(t), \qquad v\in[0,1].\]

The problem with this standard parametrisation of ruled surfaces in 
the rational case is that it is no longer of degree $n$ in $t$, since 
the denominators of the parametrisations $c(t)$ and $d(t)$ are 
different in general and hence $b(t,v)$ would be of degree $2n$.

A way of taking into account that we are dealing with rational
parametrisations would be considering polynomial parametrisations in
$\mathbb{R}^4$, \[\mathbf{p}(t)=(w(t),p(t)), \qquad
\mathbf{q}(t)=(\omega(t),q(t)).\] 

We can consider the parametrisation for a polynomial ruled surface in 
$\mathbb{R}^4$ and project back to $\mathbb{R}^3$,
\begin{equation}\label{ratpar}
\tilde 
b(t,\tilde v)=\frac{(1-\tilde v)p(t)+\tilde vq(t)}{(1-\tilde v)w(t)+
\tilde v\omega(t)},\qquad \tilde v\in[0,1],
\end{equation}
which is explicitly of degree $n$ in $t$. 

Both parametrisations are related by a change of parameters
\[v=\frac{\tilde v\omega(t) }{(1-\tilde v)w(t)+\tilde v \omega(t)}.\]

From now on we use (\ref{ratpar}) as our standard parametrisation for 
rational ruled surfaces, omitting the tilde for $v$ and $b$.

This result is useful for interpreting the functions $\Lambda$, $M$, 
$\sigma$:

Except for the case of cylindrical developable surfaces, another way 
of expressing that the tangent plane to a developable surface is the same at 
all points on the same ruling \cite{struik} is the requirement of the existence of two functions 
$\lambda(t)$, $\mu(t)$ such that
\begin{equation}\label{decomp}c'(t)=\lambda(t)\textbf{w}(t)+\mu(t)\mathbf{w}'(t),\qquad 
\mathbf{w}(t):=d(t)-c(t).\end{equation}

We can use the previous result on parametrisations of ruled surfaces 
and rewrite this condition for parametrisations 
in $\mathbb{R}^{4}$ for rational ruled surfaces in $\mathbb{R}^{3}$,
\begin{equation}\label{ratstruik}
\mathbf{p}'(t)=\lambda(t)\textbf{W}(t)+\mu(t)\mathbf{W}'(t)+ \nu(t)
\textbf{p}(t),\qquad 
\mathbf{W}(t):=\mathbf{q}(t)-\mathbf{p}(t),\end{equation}
allowing for an extra term along $\mathbf{p}(t)$ which vanishes on 
proyecting back from $\mathbb{R}^{4}$ to $\mathbb{R}^{3}$.

This term may be removed by the introduction of a suitable global 
factor $f(t)$, 
\[ \mathbf{\tilde p}(t)=f(t) \mathbf{p}(t), \qquad
 \mathbf{\tilde q}(t)=f(t) \mathbf{q}(t), \]
\begin{eqnarray*}\mathbf{\tilde p}'(t)&=&f(t) \mathbf{p}'(t)+f'(t) \mathbf{p}(t)
=
\left(\lambda(t)-\mu(t)\frac{f'(t)}{f(t)}\right)
\mathbf{\tilde W}(t)+\mu(t)\mathbf{\tilde W}'(t)+\left(\nu(t)+\frac{f'(t)}{f(t)}
\right)\mathbf{\tilde p}(t)\\&=&
\tilde\lambda(t)\mathbf{\tilde W}(t)+\tilde\mu(t)\mathbf{\tilde W}'(t)+ 
\tilde\nu(t)
\mathbf{\tilde p}(t),\end{eqnarray*}
from which we can read the terms of the new decomposition,
\[\tilde \lambda(t)=\lambda(t)-\mu(t)\frac{f'(t)}{f(t)},\quad
\tilde \mu(t)=\mu(t),\quad \tilde \nu(t)=\nu(t)+\frac{f'(t)}{f(t)},\]
and infer that the $\nu$ term can be cancelled by choosing 
\[f(t)=e^{-\int \nu(t)\,dt}.\]

We can relate these functions $\lambda,\mu,\nu$ with the ones we have 
introduced in Theorem~\ref{ratheo}, using blossom expressions for 
$\mathbf{p}$, $\mathbf{p}'$ and $\mathbf{q}$,
\begin{eqnarray}\label{blossoming}&&
\textbf{p}(t)=(1-t)\mathbf{p}[t^{<n-1>},0]+t\mathbf{p}[t^{<n-1>},1],
\nonumber\\&& \textbf{q}(t)=(1-t)\mathbf{q}[t^{<n-1>},0]+t\mathbf{q}[t^{<n-1>},1],
\nonumber\\&&
\mathbf{p}'(t)=n\mathbf{p}[t^{<n-1>},1]-n\mathbf{p}[t^{<n-1>},0],
\nonumber\\&&
\mathbf{q}'(t)=n\mathbf{q}[t^{<n-1>},1]-n\mathbf{q}[t^{<n-1>},0],
\end{eqnarray}
and grouping terms in (\ref{ratstruik}),
\begin{eqnarray*}&&\mathbf{p}[t^{<n-1>},1]\left(n(1+\mu(t))+
t(\lambda(t)-\nu(t))\right)\\&+
&\mathbf{p}[t^{<n-1>},0]\left((1-t)(\lambda(t)-\nu(t))
-n(1+\mu(t))\right)
\\&=&
\mathbf{q}[t^{<n-1>},1]\left(\lambda(t)t+n\mu(t)\right)+
\mathbf{q}[t^{<n-1>},0]\left(\lambda(t)(1-t)-n\mu(t)\right)
\end{eqnarray*}
we read from Theorem~\ref{ratheo}:
\begin{corollary}
For a rational developable surface bounded by rational curves of
degree $n$, the functions $\Lambda$, $M$, $\sigma$ in 
Theorem~\ref{ratheo} are given by
\[
\Lambda(t)=\frac{n\left(\mu(t)+1\right)+\left(\lambda(t)-\nu(t)\right)t}{\lambda(t)-\nu(t)},
\ 
M(t)=\frac{\lambda(t)t+n\mu(t)}{\lambda(t)}, \  
\sigma(t)=\frac{\lambda(t)}{\lambda(t)-\nu(t)},
\]
\[
\lambda(t)=\frac{n\sigma(t)}{\Lambda(t)-\sigma(t)M(t)+t\left(\sigma(t)-1\right)},
\ 
\mu(t)=\frac{\sigma(t)\left(M(t)-t\right)}{\Lambda(t)-\sigma(t)M(t)+t\left(\sigma(t)-1\right)},
 \]\[  
\nu(t)=\frac{n\left(\sigma(t)-1\right)}{\Lambda(t)-\sigma(t)M(t)+t\left(\sigma(t)-1\right)},
\]
in terms of the ones in the expansion (\ref{ratstruik}) for
parametrisations $\mathbf{p}(t)$, $\mathbf{q}(t)$ in $\mathbb{R}^{4}$
of the rational curves.
\end{corollary}

In this sense, Theorem~\ref{ratheo} just expresses an alternative way
of writing the rational version (\ref{ratstruik}) of the standard
decomposition (\ref{decomp}) for rational B\'ezier curves.

As expected, when $\nu\equiv 0$, no term along the projection 
direction appears and then $\sigma\equiv 1$. In this case, we 
recover the results for B\'ezier developable surfaces 
\cite{leonardo-bezier}.

These relations are useful for linking results in both formalisms 
for rational developable surfaces.

\section{Features of rational developable surfaces}

We check now how most common operations with rational curves affect 
rational developables surfaces:

\begin{itemize}
\item Multiplication of weights by a constant: If we multiply the
weights of a rational curve by a constant, the parametrisation does 
not change. 

If we multiply the list of weights of both bounding rational curves
respectively by constants $\alpha$, $\beta$, so that the new lists are
$\{\alpha w_{0},\ldots,\alpha w_{n}\}$, $\{\beta
\omega_{0},\ldots,\beta\omega_{n}\}$, 
\[\frac{\mathbf{p}[t^{<n-1>},\Lambda(t)]}{\alpha}=
\sigma(t)\frac{\mathbf{q}[t^{<n-1>},M(t)]}{\beta},\] 
it is clear that the functions
$\Lambda(t)$, $M(t)$ do not change but $\sigma(t)$ changes to 
$\alpha\sigma(t)/\beta$.

\item Reparametrisation under M\"obius transformations 
\cite{patterson,farin}: 
M\"obius transformations of the interval $[0,1]$ onto itself,
\begin{eqnarray}\label{reparmobius}
    t(u)=\frac{u}{(1-b)u+b},\quad u\in[0,1],
\end{eqnarray}
are equivalent to a change of the lists of weights 
$\{\tilde w_{0},\ldots,\tilde w_{n}\}$, 
$\{\tilde \omega_{0},\ldots,\tilde \omega_{n}\}$, $\tilde 
w_{i}=b^{n-i}w_{i}$, $\tilde \omega_{i}=b^{n-i}\omega_{i}$, 
$i=0,\ldots,n$,  while 
keeping the same control polygons for the curves.

Since we may relate both polar forms by
\begin{eqnarray*}
\frac{\mathbf{\tilde p}[u^{<n-1>},\tilde 
\Lambda(u)]}{\left((1-b)u+b\right)^{n-1}(1-b)\tilde \Lambda(u)+b}&=&
\mathbf{p}\left[\left.\frac{u}{(1-b)u+b}\right.^{<n-1>},
\frac{\tilde\Lambda(u)}{(1-b)\tilde\Lambda(u)+b}\right]\\&=&
\mathbf{p}\left[t(u)^{<n-1>},\frac{\tilde\Lambda(u)}{(1-b)\tilde\Lambda(u)+b}\right],
\end{eqnarray*}
we can write the developabity condition in terms of $\mathbf{\tilde p}$ and 
$\mathbf{\tilde q}$,
\[\mathbf{p}[u^{<n-1>},\tilde\Lambda(u)]=
\tilde\sigma(u)\mathbf{q}[u^{<n-1>},\tilde M(u)],\] 
for some new functions $\tilde\Lambda(u)$, $\tilde M(u)$, $\tilde 
\sigma(u)$, which we may relate to the previous ones,
\begin{eqnarray*}&&
\left((1-b)\tilde \Lambda(u)+b\right) 
\mathbf{p}\left[t(u)^{<n-1>},
\frac{\tilde\Lambda(u)}{(1-b)\tilde\Lambda(u)+b}\right]=
\frac{\mathbf{\tilde p}[u^{<n-1>},\tilde \Lambda(u)]}
{\left((1-b)u+b\right)^{n-1}}
\\&&=\tilde\sigma(u)\frac{\mathbf{\tilde q}[u^{<n-1>},\tilde 
M(u)]}{\left((1-b)u+b\right)^{n-1}}=\tilde\sigma(u)
\left((1-b)\tilde M(u)+b\right)
\mathbf{q}\left[t(u)^{<n-1>},
\frac{\tilde M(u)}{(1-b)\tilde M(u)+b}\right]
\end{eqnarray*}
and comparing with the developability condition in terms of 
$\mathbf{p}$ and $\mathbf{q}$, we get 
\[\Lambda(t(u))=\frac{\tilde\Lambda(u)}{(1-b)\tilde\Lambda(u)+b}, 
\quad M(t(u))=\frac{\tilde M(u)}{(1-b)\tilde M(u)+b}, \quad 
\sigma(t(u))=\tilde\sigma(u)
\frac{(1-b)\tilde M(u)+b }{(1-b)\tilde \Lambda(u)+b}, \]
\begin{equation}\label{mobius}
\tilde\Lambda(u)=\frac{b 
\Lambda(t(u))}{1+(b-1)\Lambda(t(u))},\quad \tilde M(u)=\frac{b 
 M(t(u))}{1+(b-1) M(t(u))},\quad \tilde\sigma(u)=\sigma(t(u))
 \frac{1+(b-1) M(t(u))}{1+(b-1)\Lambda(t(u))}.\end{equation}

We notice that in both cases a non-trivial $\sigma$ arises even if in 
the original parametrisation $\sigma$ is one.

\item Degree elevation: We may formally increase the degree of the
parametrisations by multiplicating $\mathbf{p}$ and $\mathbf{q}$ by
respective factors $f(t)$, $g(t)$ of degree one, which cancel out on
projecting to $\mathbb{R}^{3}$.  The new parametrisations are $
\mathbf{\tilde p}=f\mathbf{p}$ and $\mathbf{\tilde q}=g\mathbf{q}$ of degree
$n+1$.

In order to compute the functions $\tilde \Lambda$, $\tilde M$ and 
$\tilde \sigma$ for the degree-elevated parametrisations we compute, 
developing the blossom expressions,
\begin{eqnarray*}
\mathbf{\tilde p}[t^{<n>},\tilde\Lambda(t)]&=&\frac{nf(t)\mathbf{p}[t^{<n-1>},\tilde\Lambda(t)]+
f(\tilde\Lambda(t))\mathbf{p}[t^{<n>}]}{n+1}\\&=&
\frac{n(1-\tilde\Lambda(t))f(t)+f(\tilde\Lambda(t))(1-t)}{n+1}\mathbf{p}^{n-1}_{0}(t)
\\&+&
\frac{n\tilde\Lambda(t)f(t)+f(\tilde\Lambda(t))t}{n+1}\mathbf{p}^{n-1}_{1}(t),
\end{eqnarray*}
\begin{eqnarray*}
\mathbf{\tilde q}[t^{<n>},\tilde 
M(t)]&=&\frac{ng(t)\mathbf{q}[t^{<n-1>},\tilde M(t)]+
g(\tilde M(t))\mathbf{q}[t^{<n>}]}{n+1}\\&=&
\frac{n(1-\tilde M(t))g(t)+g(\tilde M(t))(1-t)}{n+1}\mathbf{q}^{n-1}_{0}(t)
\\&+&
\frac{n\tilde M(t)g(t)+g(\tilde M(t))t}{n+1}\mathbf{q}^{n-1}_{1}(t),
\end{eqnarray*}
and comparing them with the developability condition for the original 
parametrisations
\[(1-\Lambda(t))\mathbf{p}_{0}(t)+\Lambda(t)\mathbf{p}_{1}(t)=
\sigma(t)\left((1-M(t))\mathbf{q}_{0}(t)+M(t)\mathbf{q}_{1}(t)\right),\]
we read
\begin{equation}\label{raise}
\Lambda(t)=\frac{nf(t)\tilde\Lambda(t)+f(\tilde\Lambda(t))t}{nf(t)+f(\tilde\Lambda(t))},
\quad M(t)=\frac{ng(t)\tilde M(t)+g(\tilde M(t))t}{ng(t)+g(\tilde 
M(t))}, \quad
\sigma(t)=\frac{ng(t)+g(\tilde M(t))}{nf(t)+f(\tilde 
\Lambda(t))}\tilde\sigma(t).
\end{equation}


The simplest case for degree elevation is the one with $f\equiv 1\equiv g$,
\[\Lambda(t)=\frac{n\tilde\Lambda(t)+t}{n+1},
\quad M(t)=\frac{n\tilde M(t)+t}{n+1}, \quad
\sigma(t)=\tilde\sigma(t),
\]
which is the same that was found for B\'ezier developable surfaces.

\item Modification of the lengths of the rulings: We may change the
endpoints of the rulings of a developable surface patch bounded by two
rational curves $c(t)$, $d(t)$ of degree $n$ by modifying the length of
the vector $\mathbf{q}(t)-\mathbf{p}(t)$ by a linear factor $g(t)$ as
in \cite{aumann1}, $g(t)\left(\mathbf{q}(t)-\mathbf{p}(t)\right)$.

Since all terms are of degree $n+1$ but $\mathbf{p}(t)$, we may even allow 
for degree elevation of the form $f(t)\mathbf{p}(t)$, where $f(t)$ is 
a factor of degree one,
\[\mathbf{\tilde p}(t)=f(t)\mathbf{p}(t), \qquad
\mathbf{\tilde q}(t)=f(t)\mathbf{p}(t)+g(t)\left(\mathbf{q}(t)-\mathbf{p}(t)\right),\]
parametrisations which are related by the developabilty condition,
\[\mathbf{\tilde p}[t^{<n>},\tilde\Lambda(t)]=\tilde \sigma (t)
\mathbf{\tilde q}[t^{<n>},\tilde\Lambda(t)],\] for some rational 
functions $\tilde \Lambda(t)$, $\tilde M(t)$, $\tilde \sigma(t)$.

Expanding these expressions,
\[\mathbf{\tilde p}[t^{<n>},\tilde\Lambda(t)]=
\frac{f(\tilde\Lambda(t))\mathbf{p}[t^{<n>}]+
nf(t)\mathbf{p}[t^{<n-1>},\tilde\Lambda(t)]}{n+1},\]
\begin{eqnarray*}\mathbf{\tilde q}[t^{<n>},\tilde M(t)]&=&
\frac{f(\tilde M(t))\mathbf{p}[t^{<n>}]+
nf(t)\mathbf{p}[t^{<n-1>},\tilde M(t)]}{n+1}\\&+&
\frac{g(\tilde 
M(t))\mathbf{q}[t^{<n>}]+
ng(t)\mathbf{q}[t^{<n-1>},\tilde M(t)]}{n+1}\\&-&
\frac{g(\tilde 
M(t))\mathbf{p}[t^{<n>}]+
ng(t)\mathbf{p}[t^{<n-1>},\tilde M(t)]}{n+1},\end{eqnarray*}
and comparing them with
\[\mathbf{p}[t^{<n>},\Lambda(t)]= \sigma (t)
\mathbf{q}[t^{<n>},\Lambda(t)],\]
we get the relations between both sets of functions,
\begin{eqnarray}\label{modify}
\Lambda(t)&=&\frac{f(\tilde \Lambda(t))t+nf(t)\tilde \Lambda(t)+\tilde 
\sigma(t)\left(
n\tilde M(t)\left(g(t)-f(t)\right)
+t\left(g(\tilde M(t))-f(\tilde M(t))\right)\right)
}{f(\tilde \Lambda(t))+nf(t)
+\tilde 
\sigma(t)\left(g(\tilde M(t))-f(\tilde M(t))+
n\left(g(t)-f(t)\right)\right)},\nonumber\\
M(t)&=&\frac{g(\tilde M(t))t+ng(t)\tilde M(t)}{g(\tilde 
M(t))+ng(t)},\nonumber\\
\sigma(t)&=&\tilde\sigma(t)\frac{g(\tilde M(t))+ng(t)}
{f(\tilde \Lambda(t))+nf(t)
+\tilde 
\sigma(t)\left(g(\tilde M(t))-f(\tilde M(t))+
n\left(g(t)-f(t)\right)\right)}.\end{eqnarray}

In the case $f\equiv g$, we recover the expressions 
obtained for degree elevation with the same factor, since the rulings 
do not change.

Another simple case is the one with $\sigma\equiv 1\equiv f$, that 
is, with modification of the global factor just for 
$\mathbf{q}-\mathbf{p}$,
\[
\Lambda(t)=\frac{n\tilde \Lambda(t)+\left(
n\tilde M(t)\left(g(t)-1\right)
+tg(\tilde M(t))\right)}{g(\tilde M(t))+n g(t)},\]
\[M(t)=\frac{g(\tilde M(t))t+ng(t)\tilde M(t)}{g(\tilde 
M(t))+ng(t)},\qquad \sigma\tilde \equiv 1. \]

\item Knot insertion: One of the advantages of writing the 
developability condition in terms of blossoms is the invariance under 
knot insertion. That is, if we insert a new knot in the list, the 
expressions for $\Lambda$, $M$ and $\sigma$ will not change.
\end{itemize}

\section{Edge of regression}

The edge of regression is the set of points of the developable 
surface where the surface is singular. The coordinate patch $(t,v)$ 
fails at the edge, since the vectors $b_{t}$ and $b_{v}$ are 
parallel. The developable surface can be seen as the tangent surface 
to its edge of regression \cite{struik}, except for the cases of 
cylindrical and conical surfaces. 

We may compute the edge of regression of a rational developable 
surface, making use of the polynomial parametrisations of the 
bounding curves, $\mathbf{p}(t)$ and $\mathbf{q}(t)$ in 
$\mathbb{R}^{4}$. 

Before projection onto $\mathbb{R}^{3}$, the polynomial 
parametrisation would be
\[\textbf{b}(t,v)=(1-v)\mathbf{p}(t)+v\mathbf{q}(t),\]
but in this case we cannot simply require parallelism, 
$\mathbf{b}_{t}=\alpha \mathbf{b}_{v}$, of the 
derivatives
\[\mathbf{b}_{t}(t,v)=(1-v)\mathbf{p}'(t)+v\mathbf{q}'(t),
\qquad \mathbf{b}_{v}=\mathbf{q}(t)-\mathbf{p}(t),\]
but allow an additional term along $\mathbf{b}$ which vanishes on projecting to 
$\mathbb{R}^{3}$,
\[(1-v)\mathbf{p}'(t)+v\mathbf{q}'(t)=\alpha(t,v)\left(\mathbf{q}(t)-\mathbf{p}(t)\right)+
\beta(t,v)\left((1-v)\mathbf{p}(t)+v\mathbf{q}(t)\right)\]

Following (\ref{blossoming}), we may group terms in the previous 
expression,
\begin{eqnarray*}
&&\mathbf{p}[t^{<n-1>},0]\left(n(v-1)+(1-t)\left(\alpha(t,v)+ 
(v-1)\beta(t,v)\right)\right)\\&+&
\mathbf{p}[t^{<n-1>},1]\left(n(1-v)+t\left(\alpha(t,v)+ 
(v-1)\beta(t,v)\right)\right)\\&=&
\mathbf{q}[t^{<n-1>},0]\left(nv+(1-t)\left(\alpha(t,v)+ 
v\beta(t,v)\right)\right)\\&+&
\mathbf{q}[t^{<n-1>},1]\left(-nv+t\left(\alpha(t,v)+ 
v\beta(t,v)\right)\right),
\end{eqnarray*}
and compare them with the ones in Theorem~\ref{ratheo} to yield
\[\Lambda(t)=t+\frac{n(1-v)}{\alpha(t,v)+(v-1)\beta(t,v)},
\quad M(t)=t-\frac{nv}{\alpha(t,v)+v\beta(t,v)},\]
\[\sigma(t)=\frac{\alpha(t,v)+v\beta(t,v)}{\alpha(t,v)+(v-1)\beta(t,v)},
\]
from which we can eliminate $\alpha$ and $\beta$, which happen not to 
depend on $v$,
\[\alpha(t)=\frac{n\sigma(t)\left(\Lambda(t)-M(t)\right)}
{\left(\Lambda(t)-t+\sigma(t)\left(t-M(t)\right)\right)^{2}},\quad
\beta(t)=\frac{n\left(\sigma(t)-1\right)}
{\Lambda(t)-t+\sigma(t)\left(t-M(t)\right)},\]
and we obtain a parametrisation of the edge of regression:
\begin{corollary}A rational developable surface with bounding curves 
$c(t)$, $d(t)$ satisfying
\[\mathbf{p}[t^{<n-1>},\Lambda(t)]=\sigma(t)\mathbf{q}[t^{<n-1>},M(t)],\]
for some rational functions $\Lambda$, $M$, $\sigma$ has a rational edge of 
regression parametrised in $\mathbb{R}^{4}$ by
\begin{equation}\textbf{r}(t)=(1-v(t))\mathbf{p}(t)+v(t)\mathbf{q}(t),\qquad v(t)=\frac{\sigma(t)\left(t-M(t)\right)}{\Lambda(t)-t+\sigma(t)\left(t-M(t)\right)}.\end{equation}
\end{corollary}

%
%
 We notice that 
in the case of constant $\Lambda$, $M$, $\sigma$, the edge of 
regression is a rational curve of degree $n+1$, since the parametric 
equation $v(t)$ is of degree one.

Hence, developable surfaces bounded by rational curves of degree $n$
with constant functions $\Lambda$, $M$, $\sigma$ have rational edges
of regression of degree $n+1$.


We may check if the converse is also true. That is, if all rational 
developable surfaces with rational edge of regression of degree $n$ 
have surface patches of degree $(n-1,1)$ with constant functions 
$\Lambda$, $M$, $\sigma$:

The simplest generic developable surface is the tangent surface to a 
curve $\textbf{r}(t)$ of degree $n$, which is the edge of regression of the 
developable surface, 
\[\mathbf{b}(t,v)=\mathbf{r}(t)+vf(t)\mathbf{r}'(t),\]
which is a rational developable surface patch of degree $(n,1)$ 
provided that the factor $f$ is linear, $f(t)=at+b$.

According to (\ref{ratstruik}),  $\mathbf{r}'(t)=\mathbf{W}(t)/f(t)$ 
and hence $\lambda(t)=1/f(t)$, $\mu(t)=0$, $\nu(t)=0$, so that 
\[\Lambda(t)=t+nf(t),\qquad M(t)=t,\qquad \sigma\equiv 1.\]

We may choose another surface patch bounded by two curves of degree
$n-1$ on the developable surface, since it is clear that if we take
$v=1$, $a=-1/n$ on the parametrisation, all resulting curves,
depending on $b$, are of degree $n-1$, since we have removed the
leading term in $t$ in their parametrisations.

With the following choice of the free parameter $b$,
\[\mathbf{p}(t)=\mathbf{r}(t)+\frac{M-t}{n}\mathbf{r}'(t),\qquad
\mathbf{q}(t)=\mathbf{r}(t)+\frac{\Lambda-t}{n}\mathbf{r}'(t),\]
it is easy to check that the developable surface patch bounded by the
respective curves $c(t)$ and $d(t)$ of degree $n-1$ has constant
functions $\Lambda$, $M$, $\sigma=1$:

\begin{theorem}The set of developable surfaces with patches generated by two 
rational curves $c(t)$, $d(t)$ of degree $n$, with ruling generators 
$\mathbf{q}(t)-\mathbf{p}(t)$ also of degree $n$ and blossoms related by
\[\mathbf{p}[t^{<n-1>},\Lambda]=\mathbf{q}[t^{<n-1>},M],\] with
constant $\Lambda$, $M$, $\sigma=1$, is the set of tangent surfaces to rational
curves of degree $n+1$.\end{theorem}

That is, we may generate the rational developable surfaces with
constant $\Lambda$, $M$, $\sigma=1$ patches and then adapt these 
patches.

The proof is simple, writing (\ref{ratstruik}) in our case,
\[\mathbf{W}(t)=\mathbf{q}(t)-\mathbf{p}(t)=
\frac{\Lambda-M}{n}\mathbf{r}'(t),\qquad \mathbf{W}'(t)=
\frac{\Lambda-M}{n}\mathbf{r}''(t),\]
\[\lambda(t)\mathbf{W}(t)+\mu(t)\mathbf{W}'(t)=
\mathbf{p}'(t)=\mathbf{r}'(t)-\frac{\mathbf{r}'(t)}{n}+\frac{M-t}{n}\mathbf{r}''(t)
=\frac{n-1}{\Lambda-M}\mathbf{W}(t)+\frac{M-t}{\Lambda-M}\mathbf{W}'(t),\]
from which we read the functions $\lambda(t)$, $\mu(t)$, $\nu(t)$,
\[\lambda(t)=\frac{n-1}{\Lambda-M},\qquad \mu(t)=\frac{M-t}{\Lambda-M}
,\qquad \nu(t)=0,\]
which correspond to the right values of $\Lambda$, $M$, $\sigma$, according to 
Corollary~2, for $c(t)$ of degree $n-1$.

%
%
%
%
%

\section{The constant $\Lambda$, $M$, $\sigma$ case\label{constante}}

The simplest case which can be considered is the one with constant
coefficients $\Lambda$, $M$, $\sigma$,
\[\mathbf{p}[t^{<n-1>},\Lambda]=\sigma\mathbf{q}[t^{<n-1>},M].\] 


This expression states the equality of two $(n-1)$-atic forms, which 
is equivalent to the equality of the respective symmetric 
$(n-1)$-affine forms, since the correspondence between blossoms and 
parametrizations is one-to-one,
\begin{equation}
\mathbf{p}[t_{1},\ldots,t_{n-1},\Lambda]=\sigma\mathbf{q}[t_{1},\ldots,t_{n-1},M]\
.\end{equation}

We may draw information about the control net applying it to sequences
of zeros and ones, taking into account that the vertices are recovered
as \[\mathbf{p}_{j}=\mathbf{p}[0^{<n-j>},1^{<j>}],\]
\[(1-\Lambda)\mathbf{p}_{j}+\Lambda
\mathbf{p}_{j+1}=(1-M)\sigma\mathbf{q}_{j}+M \sigma\mathbf{q}_{j+1},\quad
j=0,\ldots,n-1,\] stating that the cells of the control net of the
surface in $\mathbb{R}^4$ are planar and share the same linear 
combination of vertices.

Projecting back into affine space, we get the relation between 
weights and vertices of the control net,
\[(1-\Lambda)w_{j}+\Lambda w_{j+1}=
(1-M)\sigma\omega_{j}+M\sigma\omega_{j+1},\]\[(1-\Lambda)w_{j}c_{j}+
\Lambda w_{j+1}c_{j+1}=(1-M)\sigma\omega_{j}d_{j}+M\sigma\omega_{j+1}d_{j+1}.\]

This can be considered the natural 
generalization of Aumann's result for B\'ezier developable surfaces 
to the rational case \cite{aumann}, though in that paper the key issue was the use 
of an affine transformation between adjacent cells of the control net 
of the surface.

\section{Example}

We construct the rational developable surface bounded by a curve $c(t)$
with control polygon $\{(0, 0, 0), (3, 3, 0), (4, 3, 0), (5,0,0)\}$ 
and weights $\{1,1,3/5,5/6\}$ and a curve $d(t)$ with 
$d_{0}=(0, 0, 2)$, $d_1=(2, 2, 3)$, $\omega_{0}=1$, $\omega_{1}=1$.
\begin{figure}
\begin{center}
    \includegraphics[height=0.2\textheight]{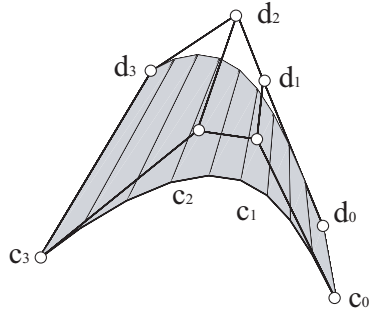}
\end{center}
\caption{Developable surface bounded by rational cubic curves}
\label{example}\end{figure}
We complete the control polygon,
\[\{(0, 0, 2), (2, 2, 3), (3/2, 21/22, 135/22), (293/172,
-189/172, 1215/172)\}, \] and list of weights of $d(t)$,
\[ \{1,1, 11/15, 43/45\},\]
building a patch with constant $\Lambda,M,\sigma$ as in
Section~\ref{constante}, for which $\Lambda=-4/3$, $M=-2$, $\sigma=1$.  The
resulting developable surface is shown in Fig.~\ref{example}.

The edge of regression is the curve parametrised by
\[r(t)=(1-v(t))c(t)+v(t)d(t),\qquad v(t)=
\frac{\sigma(t-M)}{\Lambda-t+\sigma(t-M)}=\frac{3}{2}t+3,\]
according to corollary~3.

If we perform a M\"obius transformation (\ref{reparmobius}) with 
$b=2$ on both bounding curves, so that the respective new sets of 
weights are $\{8,4,6/5,5/6\}$ and $\{8,4,22/15,43/45\}$, 
the new parametrisation for the 
developable surface has new constants given by (\ref{mobius}),
\[
\tilde\Lambda=\frac{b 
\Lambda}{1+(b-1)\Lambda}=8,\quad \tilde M=\frac{b 
 M}{1+(b-1) M}=4,\quad \tilde\sigma=\sigma
 \frac{1+(b-1) M}{1+(b-1)\Lambda}=3.\]
 
\begin{figure}
\begin{center}
    \includegraphics[height=0.2\textheight]{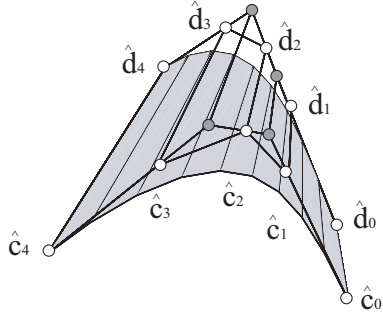}
\end{center}
\caption{Degree-elevated developable surface patch}
\label{ratelev}\end{figure}

If we raise the degree of both curves in the usual way, with $f\equiv 
1\equiv g$, the control polygons of the curves change to (see 
Fig.~\ref{ratelev})\[\{( 0, 0, 0), 
(9/4, 9/4, 0), (27/8, 3, 0), (341/79, 162/79, 0), (5, 0, 0)\},\]\[ 
\{(0, 0, 2), (3/2, 3/2, 11/4), (93/52, 81/52, 225/52), 
(887/568, 189/568, 3645/568), (293/172, -189/172, 1215/172)\},\] and 
the lists of weights to $\{1, 1, 4/5, 79/120, 5/6\}$, $\{1, 1, 13/15, 
71/90, 43/45\}$. According to (\ref{raise}), the new functions are
\[\tilde \Lambda(u)=-\frac{u}{3}-\frac{16}{9} 
,\qquad \tilde M(u)=-\frac{u}{3}-\frac{8}{3},\qquad
\tilde\sigma(u)=1.\]
 
\begin{figure}
\begin{center}
    \includegraphics[height=0.2\textheight]{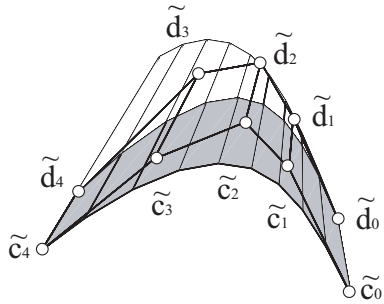}
\end{center}
\caption{Trimmed developable surface patch}
\label{ratshort}\end{figure}

We may shorten the lengths of the rulings of our patch, since the 
last ruling is much larger than the first one. For instance, we may 
move the boundary of the patch from $d(t)$ to $\tilde 
d(t)=c(t)+g(t)(d(t)-c(t))$. If we take $g(t)=1-3t/4$, we get, 
according to (\ref{modify}),
\[\tilde\Lambda(t)=\frac{1}{9}\frac{9t^2-3t-32}{1-2t}, \quad 
M(t)=\frac{1}{3}\frac{7t-16}{1-2t},\quad \sigma(t)=1,\]
since we have not introduced a global factor $f(t)$ for $c(t)$. The 
result may be seen in Fig.~\ref{ratshort}. 

\begin{figure}
\begin{center}
    \includegraphics[height=0.2\textheight]{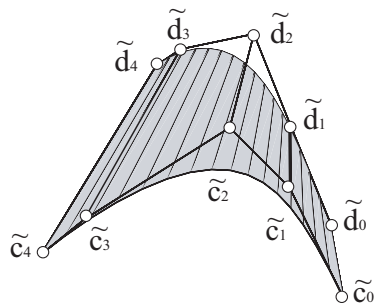}
\end{center}
\caption{Knot-inserted developable surface patch}
\label{ratspline}\end{figure}

Finally, we consider our surface patch as a spline patch of one piece
with trivial lists of knots, $\{0,0,0,1,1,1\}$, and $\{0,1\}$. If we 
insert a new knot $t=2/3$ (see Fig.~\ref{ratspline}), $\Lambda$, $M$, 
$\sigma$ do not change.

\section{Conclusions}

In this paper a new characterization of rational and NURBS developable
surfaces in terms of blossoms of their bounding curves has been
produced.  This is useful for CAD purposes, since the bounding curves
are described in terms of their control points, weights and knots.  As
a consequence, a way of constructing generic rational developable
surfaces has been shown, using linear relations between the control
points and weights of the cells of the control net of the surface.
This construction is compatible with algorithms based on blossoms and
allows easily elevation of degree.  The edge of regression of the
developable surface has a simple expression in terms of the parameters
used for deriving the construction.


\section*{Acknowledgments}
 
This work is partially supported by the Spanish Ministerio de
Econom\'\i a y Competitividad through research grant TRA2015-67788-P.


\end{document}